\documentclass[sigconf, nonacm]{acmart}
\AtBeginDocument{%
  }
\setcopyright{acmcopyright}
\copyrightyear{2022}
\acmYear{2022}
\setcopyright{rightsretained}
\acmConference[UIST '22 Adjunct]{The Adjunct Publication of the 35th Annual ACM Symposium on User Interface Software and Technology}{October 29-November 2, 2022}{Bend, OR, USA}
\acmBooktitle{The Adjunct Publication of the 35th Annual ACM Symposium on User Interface Software and Technology (UIST '22 Adjunct), October 29-November 2, 2022, Bend, OR, USA}
\acmDOI{10.1145/3526114.3558697}
\acmISBN{978-1-4503-9321-8/22/10}
\begin{document}
\title{ARfy: A Pipeline for Adapting 3D Scenes to Augmented Reality}
\author{Arthur Caetano}
\orcid{0000-0003-0207-5471}
\affiliation{%
  \institution{University of California, Santa Barbara}
  \city{Santa Barbara}
  \state{California}
  \country{USA}
  \postcode{93106}
}
\author{Misha Sra}
\affiliation{%
  \institution{University of California, Santa Barbara}
  \city{Santa Barbara}
  \state{California}
  \country{USA}
  \postcode{93106}
}
\renewcommand{\shortauthors}{Caetano et al.}
\begin{abstract}
Virtual content placement in physical scenes is a crucial aspect of augmented reality (AR). This task is particularly challenging when the virtual elements must adapt to multiple target physical environments unknown during development. AR authors use strategies such as manual placement performed by end-users, automated placement powered by author-defined constraints, and procedural content generation to adapt virtual content to physical spaces. Although effective, these options require human effort or annotated virtual assets. As an alternative, we present ARfy, a pipeline to support the adaptive placement of virtual content from pre-existing 3D scenes in arbitrary physical spaces. ARfy does not require intervention by end-users or asset annotation by AR authors. We demonstrate the pipeline capabilities using simulations on a publicly available indoor space dataset. ARfy makes any generic 3D scene automatically AR-ready and provides evaluation tools to facilitate future research on adaptive virtual content placement.
\end{abstract}



\maketitle

\section{Introduction}

Augmented reality (AR) allows users to view and interact with virtual elements integrated with the real world. The problem of associating virtual elements to real-world coordinates is referred to as registration\cite{azuma1997survey} and view management\cite{bell2001viewmanagement}. In this work, we refer to both as \textit{AR placement}. It is a desirable property of AR applications to have \textit{adaptive} AR placement, to provide a consistent experience in different physical environments.

Prior work has explored various solutions to the AR placement problem. In many consumer applications, users place virtual elements in their physical environment manually supported by planar detection which creates tracked meshes where virtual elements are anchored, achieving perspective-corrected visualization. Researchers have proposed improvements to this kind of interaction \cite{nuernberger2016snaptoreality}. Users can also place virtual content in a physical location using AR markers. Even though this approach has become less popular with the growth of computer-vision-based surface detection methods, it is still an effective alternative for scenarios where tracking is unstable due to clutter or reduced lighting \cite{garrido2014aruco}.

Another category of solutions uses dynamic placement algorithms to avoid delegating AR placement to end-users. These algorithms find placements that satisfy constraints inferred from the physical environment and defined by authors. The constraints often present themselves in the form of virtual asset annotations.  Recent work has explored different types of constraints such as semantic, utilitarian, affordance related\cite{qian2022scalar, cheng2021semanticadapt, lindlbauer2019context}, geometric\cite{unity2022mars, hettiarachchi2016annexing}, spatial\cite{gal2014flare, marwecki2018scenograph, singh2021story}, cognitive, and geospatial\cite{niantic2022lightship, burkard2020geospatial}.

Yet another alternative for AR placement is to fit the virtual elements to a specific physical location. This approach allows authors to exploit site-specific geometry and affordances for tailor-made AR experiences \cite{lin2020architect}. Access to space geometry also allows seamlessly aligned AR placement for maintaining the illusion of realism \cite{azuma1997survey}. Procedural content generation can produce virtual elements with a high degree of variability for such environments \cite{sra2018oasis,cheng2019vroamer, yang2019dreamwalker}. Scene-specific methods have also explored replacing real-world objects using inpainting and augmentations \cite{kari2021transformr, litany2017asist}. 

In this work we present ARfy, a pipeline to adapt existing 3D scenes to arbitrary physical environments, as a novel solution to the AR placement problem. Unlike prior methods, ARfy does not require modifications on the original assets and assumes no prior knowledge of the target physical location. The pipeline uses point cloud registration to estimate a 3D transformation from the point cloud extracted from the virtual scene to the point cloud from the physical space \cite{cheng2018registration}. Prior end-user applications have used this approach to perform AR placement of CAD models with 1:1 correspondence to physical objects \cite{wu2018augmented, mahmood2020bim}. In contrast, our pipeline is designed to fit into the AR development process. It allows any virtual model to align with any physical space model. Our main contributions are as follows:
\begin{enumerate}
  \item ARfy allows authors to convert non-AR 3D scenes into adaptive AR experiences without modifying the original 3D assets.
  \item It improves the user experience by relieving end users from the manual AR placement task.
  \item It supports researchers and AR authors with testing and visualization tools to evaluate placement algorithms based on point cloud data.
\end{enumerate}

\section{Pipeline Design}

\begin{figure}
\centering
\includegraphics[width=8cm]{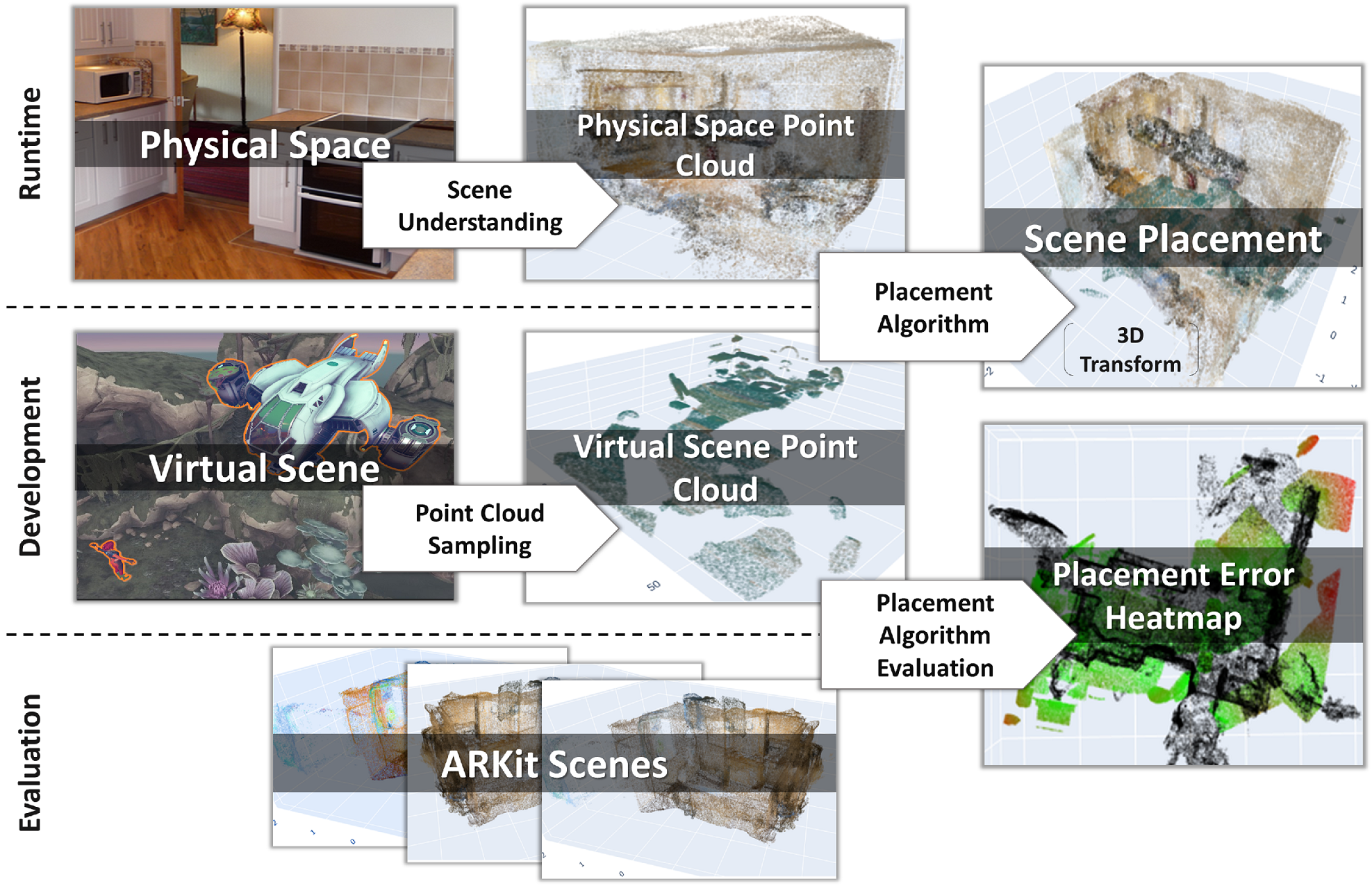}
\setlength{\belowcaptionskip}{-15pt}
\caption{\textbf{Runtime}: extracting point cloud from physical space to align virtual scene. \textbf{Development}: extracting point cloud from virtual scene to support runtime and evaluation. \textbf{Evaluation}: visualizing placement error of the virtual scene in the test dataset as a heatmap.}
\label{fig:arfy}
\end{figure}

The central idea in the ARfy pipeline is representing a 3D scene as point cloud data. We use this data structure in three stages of the proposed pipeline, shown in Figure \ref{fig:arfy}. For easier adoption, we built ARfy as a combination of Python Web API and Unity plugin\cite{python2022lang, unity2022engine}. It works on both mobile and head-mounted AR devices.

\subsection{Development Stage}
\label{subsec:development}

The first stage is to convert a pre-existing 3D scene into a point cloud (Figure \ref{fig:arfy} development).  We accomplished this by using a Unity component to sample in two ways. It can select points from the surface of the 3D meshes, resulting in a point cloud covering the scene. Alternatively, it can take the center point of support planes, i.e., the lowest XZ face of an object's bounding box. These two procedures enable more flexibility in the result, as discussed in section \ref{sec: discussion}. Authors can use the Unity layer system to assign layers to the sampling methods or mark them to be ignored by the sampler.

\subsection{Evaluation Stage}

Due to the high variability of physical environments encountered in the real world, it is challenging to predict the quality of the AR placement in advance. ARfy includes an automated test tool to tackle this issue. Our test tool executes a point cloud-based placement algorithm with a virtual scene as input and measures the placement error over a physical space dataset. The placement quality feedback is presented to the author as a heatmap of placement error at point-level (Fig. \ref{fig:arfy} evaluation), insightful to tune the virtual scene or AR placement algorithm. Our current implementation allows authors to test their 3D on ARKit Scenes, a publicly available indoor scene point cloud dataset \cite{dehghan2021arkitscenes}.

\subsection{Runtime Stage}
\label{subsec:runtime}

Current AR platforms\cite{apple2022arkit, google2022arcore} can extract point cloud data from scanned physical scenes. The AR placement algorithm receives as inputs the the scanned point cloud data and a reference to the 3D scene point cloud, stored earlier. The AR placement is, thus, reduced to a point cloud registration problem, which has been extensively explored \cite{makadia2006autoreg,pandey2017alignment, cheng2018registration}. Given two point cloud inputs of the 3D scene and the physical location (Figure \ref{fig:arfy} runtime), it is possible to compute a 3D transformation (translation, rotation, and scaling) that minimizes an error metric between the two. We use the mean square distance between pairs of closest points from the point clouds as our error metric. The optimization takes into account two restrictions. First, rotation is constrained to the Y-axis, because we assume the virtual and real-world XZ planes must be aligned. Second, scale is defined by a single scalar value, because we assume a AR placement that preserves the aspect-ratio of the 3D scene. As a baseline, we provide a simple implementation of the Iterative Cloud Registration (ICP) algorithm \cite{besl1992icp} to compute a placement transformation.

\section{Demonstration}
\label{sec:demo}

\begin{figure}
\includegraphics[height=4cm, width=8cm]{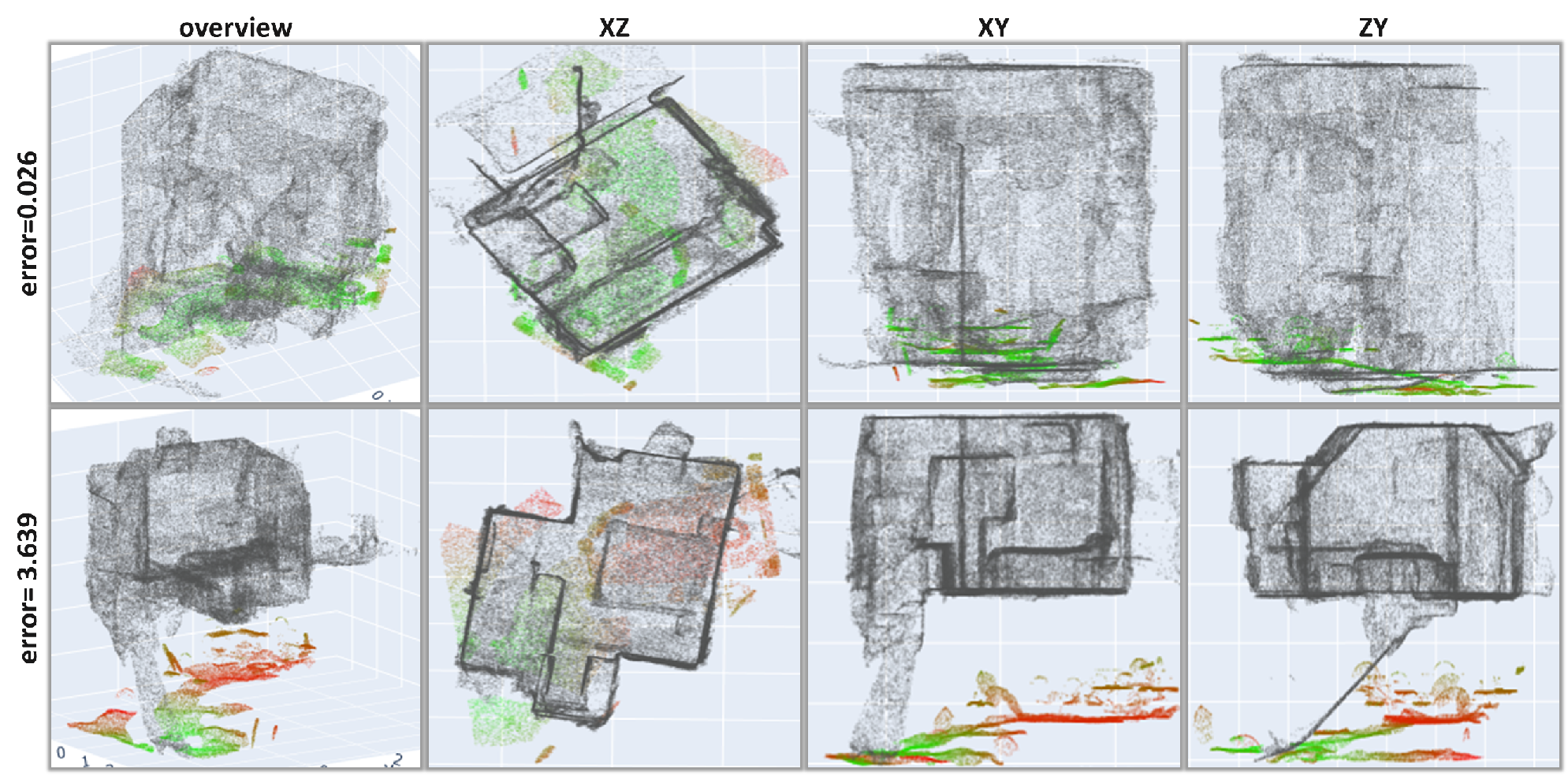}
\setlength{\belowcaptionskip}{-15pt}
\caption{\textbf{Top}: best alignment with error 0.026 on video id 47332911. \textbf{Bottom}: worst alignment with error 3.639 on video id 45261190.}
\label{fig:demo}
\end{figure}

To demonstrate the ARfy pipeline, we adapt the 3D scene from Unity Game Kit\cite{unity2021gamekit} to the real-world indoor spaces from the ARKit Scenes dataset \cite{dehghan2021arkitscenes}. We randomly sampled point clouds of size 1000 from the original dense point clouds. We adopted the mean square distance between pairs of closest points as our error metric. The scene with video id 47332911 had the best placement, with an error of 0.026. The scene with video id 45261190  had the worst placement, with an error of 3.639. Results are shown in Figure \ref{fig:demo}. Results are directly comparable because they normalize error metrics by physical space and use the same 3D scene as the test instance.

\section{Discussion, Limitations and Future Work}
\label{sec: discussion}

As demonstrated, ARfy can solve the AR placement problem satisfactorily in a wide range of physical locations. We observed that the most challenging physical spaces for adapting the tested 3D scene were spaces with mirrors and multiple floors. As expected, ARfy adapted the virtual scene to empty rectangular rooms with higher precision. Future work can explore other point cloud registration and optimization methods. Detecting and removing outliers could also lead to improvements in placement quality.

ARfy brings a new capability to the AR ecosystem and can integrate with other tools for design of end-to-end Adaptive AR systems. Our current pipeline depends on underlying AR platforms \cite{apple2022arkit,google2022arcore, unity2022engine} for tracking, point cloud extraction, rendering in world-coordinates, physics, etc. Our solution can coexist with other AR placement strategies by producing an initial coarse placement that can be manually adjusted or refined incrementally. 

Although ARfy focuses on automatic adaptive AR placement, it is possible to alter results by changing the input parameters. For example, to create the effect of 3D elements laying directly on top of real-world surfaces \cite{microsoft16conker}, authors can assign relevant foreground elements to the support point procedure described in Section \ref{subsec:development} and ignore other layers. For better results in this setting, authors can set the flag \textit{keep-aspect-ratio} to false, only render the relevant layers, and integrate physical surface geometry into the collision detection system of a physics engine. 

While our demonstration is limited to indoor scenes from the dataset, we plan to test ARfy for outdoor physical locations in the future. To achieve AR placement for outdoor experiences, we can use the scale factor to be one over the camera height. Additionally, pre-scanned point clouds of outdoor spaces combined with geolocation could serve as alternatives to the point clouds extracted by the user's AR device at runtime.

\section{Conclusion}
In this work, we presented ARfy, a pipeline that automatically adapts existing 3D scenes into AR by placing them in different physical locations without requiring asset annotation or manual placement. The demonstration shows evidence of the range of placement quality achieved by the current version and the feasibility of the pipeline.

\begin{acks}
We would like to thank Apurv Varshney for contributing to the problem formulation and initial discussions, and Prof. Ambuj K. Singh, who provided guidance and insightful feedback. 
\end{acks}

\bibliographystyle{ACM-Reference-Format}
\bibliography{bib}

\end{document}